\documentclass[aps,twocolumn,pre,showpacs,floatfix]{revtex4}
\usepackage{graphicx}

\begin{document}

\title{{\Large Electrophoresis of a Charge-Inverted Macroion Complex: \\
Molecular Dynamics Study}}

\author{Motohiko Tanaka} \affiliation{ National Institute for Fusion 
Science, Toki 509-5292, Japan} \author{A.Yu.Grosberg} \affiliation{
Department of Physics, University of Minnesota, Minneapolis, MN 55455 }

\begin{abstract} 
We have performed molecular dynamics simulations to
study the effect of an external electric field on a macroion in
the solution of multivalent $Z:1$ salt. To obtain plausible
hydrodynamics of the medium, we explicitly make the simulation
of many neutral particles along with ions. In a weak
electric field, the macroion drifts together with the strongly
adsorbed multivalent counterions along the electric field, in
the direction proving inversion of the charge sign. The
reversed mobility of the macroion is insensitive to the external
field, and increases with salt ionic strength. The reversed
mobility takes a maximal value at intermediate counterion valence.
The motion of the macroion complex does not induce any flow of the
neutral solvent away from the macroion, which reveals screening of
hydrodynamic interactions at short distances in electrolyte
solutions. A very large electric field, comparable to the
macroion unscreened field, disrupts charge inversion by stripping
the adsorbed counterions off the macroion.
\end{abstract}


\maketitle

\section{Introduction}
\label{Sec.1}

The concept of electrostatic screening has been well known
since the work by Debye and H\"uckel of early $ 20th $ century
\cite{DH}.
In recent years, screening by strongly charged ions was found
to result in counterintuitive phenomena such as attraction between
like-charged macroions \cite{Crock,Rouzina,Linse,Michael}, and
inversion of macroion charges
\cite{deJong,Straus,Elime,Dubin,Kaba1,Grant,%
Kaba2,SafinyaDendrimers,Biomol,Mengen,Gonza,Sjost,%
Kjell,Shklovskii,NguGS,Monica,Mainz,Tanaka,Tovar,review}. Charge
inversion was studied by experiments using both colloids and
biological materials
\cite{deJong,Straus,Elime,Dubin,Kaba1,Grant,Kaba2,%
SafinyaDendrimers,Biomol}, by analytical theories
\cite{Gonza,Kjell,Shklovskii,NguGS,Monica,Tovar,review}, and by
Monte Carlo and molecular dynamics (MD) simulations
\cite{Mengen,Sjost,Mainz,Tanaka}.

Experimentally, the most direct method to observe charged colloids
and macroions is electrophoresis. This method is also the prime
candidate for the technique of observing charge inversion.
Although straightforward, this approach involves many questions
upon a closer look. For example, does the macroion drift along
with its adsorbed multivalent ions when an external electric
field is applied? How many multivalent ions are attached to the
macroion strongly enough to drift together? How is the drift
affected by the solvent viscosity and the counterflow of
monovalent ions? What happens when the field becomes very strong?
What is the field strength sufficient to disrupt the charged
complex? These are the fundamental questions necessary to address
in order to bridge theoretical concepts of charge inversion and
experimental observations.

It should be born in mind that electrophoresis in general has
quite a few delicate aspects. Some peculiar ones attracted
significant attentions recently \cite{Ajdari} (see also the review
article \cite{Viovy} and references therein). The electric field
acts not only on the macroions, but on every ion in a solution. In
many cases, this leads to effective screening of hydrodynamic
interactions which otherwise may be very significant. In the
simulations reported below, we have actually observed such
short-range screening of hydrodynamic interactions in the system
comprising of a macroion, counterions and coions
(Sec.\ref{Sec.3.2.1}).

One of the difficulties in simulating charge inversion under
electrophoresis consists in subtle interactions of a macroion with
surrounding ions and neutral solvent. A naive use of the Langevin
equation, assuming that every ion (radius $ R $) in the system is
subject to Stokesian friction $ - 6 \pi \eta R v $ and white noise
random forces that balance the friction through the
fluctuation-dissipation theorem, is not justifiable. A simple
counterexample would be two closely located spheres. Since other
particles (either ions or neutral solvent) are excluded from the
volume between spheres, neither their corresponding friction
forces nor random forces add to each other. One way of going
around this problem is to incorporate macroion-solvent
interactions using the Oseen tensor
\cite{LandauLifshitzHydrodynamics}. This is, however, not easy to
implement in numerical simulations, because the interactions
produce complicated spatial correlations among random forces.
Therefore, we address this problem by a direct approach
introducing explicit neutral particles to deal with the
macroion-salt-solvent interactions in the molecular dynamics
simulations.

The explicit simulation of the solvent molecules is of course
costly. In this paper, with the limited computational resources,
we restrict ourselves to the system with only the $Z:1$ salt,
and no $1:1$ salt. It is needless to say that in real water
solvent there is always some amount of $ 1:1 $ salt. In this sense,
our present paper demonstrates the principle that charge inversion
is a phenomenon observable by a direct electrophoresis experiment.
Further study including the $1:1$ salt will be required to compare
results with realistic systems. Here, we will specify the
deviations arising from the lack of the $1:1$ salt.
Also, we will confine ourselves to the study of a single
macroion interacting with surrounding salt ions and solvent.
%We postpone for the future the study of a more complicated case
%involving several macroions and interactions between them.

Our plan in this study is to examine electrophoresis of a
spherically shaped, uniformly charged macroion. We will
systematically measure the mobility of the drifting
macroion complex placed under an external electric field by
molecular dynamics simulations. We first show that charge
inversion does take place in a solution containing multivalent
counterions, as manifested by the inverse mobility under the weak
electric field. We then look at the dependences of the reversed
mobility on the parameters, such as the concentration of co- and
counterions and the surface charge density of the macroion.
Finally, we consider the strong electric field regime and show
that the strong field disrupts the charge-inverted macroion
complex and terminates the charge inversion phenomenon.

\section{Simulation Model and Parameters}
\label{Sec.2}

\subsection{Description of the Model}

We adopt the following model, with $ a $, $ e $, and $ m $ being
the units of length, charge and mass, respectively. (We have in
mind $ a \sim 2 $\AA \ and $ m \sim 40 $ a.m.u.) A macroion with
negative charge $ Q_{0} $ between $ -15e $ and $ -180e $ is
surrounded by $ N^{+} $ counterions of a positive charge $ Ze $
and $ N^{-} \approx 300 $ coions of a negative charge $ -e $. The
system is maintained in overall charge neutrality, $ Q_{0} +
N^{+}Ze - N^{-}e = 0 $, which determines $ N^{+} $ for a given $ Z
$. The mass of the macroion is $ M= 200 m $, and the mass of the
co- and counter-ions is $ m $. We also include $ N_{*} $ neutral
particles with mass $ m/2 $, where we note the mass of water
molecule against that of K$ {}^{+} $ or Ca$ {}^{2+} $ ions.
Approximately one neutral particle is located in every volume
element $ (1.5a)^{3} \approx (3 $\AA$)^{3} $ inside the simulation
domain, excluding the locations already occupied by the macroion
and other ions, which typically yields 8000 neutral particles.
These particles are confined in a rectangular box of size $ L $,
with periodic boundary conditions in all three directions. Most
of the runs are performed for $L=32a$, except one series of the
runs intended to test the finite size effect of the domain
(Fig.\ref{Fig.mu_L}) described in Sec.\ref{Sec.3.2.1}.

In addition to the Coulomb forces, all particles interact through
the repulsive Lennard-Jones potential $ \phi_{LJ}= 4 \varepsilon
[(\sigma/r_{ij} )^{12} - ( \sigma/r_{ij} )^{6}] $ for $ r_{ij}=
|{\bf r}_{i}-{\bf r}_{j}| \le 2^{1/6}\sigma $, and $ \phi_{LJ}= -
\varepsilon $ otherwise. Here $ {\bf r}_{i}$ is the position vector
of the $ i $-th particle, and $ \sigma $ is the sum of the radii
of two interacting particles, which are chosen as follows: radius
of the macroion, $ R_{0} $, is between $ 3a $ and $ 5a $,
counterions and coions have radius $ a $, and neutral particles $
a/2 $. We relate $ \varepsilon $ with the temperature by
$ \varepsilon = k_{B}T $, and choose $ k_{B}T = e^{2}/5 \epsilon a $
(we assume spatially homogeneous dielectric constant $ \epsilon $).
The Bjerrum length is thus $ \lambda_{B}= e^{2}/\epsilon k_{B}T=
5a $. For the parameters of this
temperature, the valence $ Z=3 $, and the number of coions $
N^{-}= 300 $, the ionic strength becomes $ n_{I}= (Z^{2} N^{+}
+N^{-})/L^{3} \sim 3.7 \times 10^{-2} a^{-3} $.

After knowing the ionic strength, one is tempted to compute the
Debye length which turns out to be $ \lambda_{D} \sim 0.45a $. We
should emphasize that this number does not have much meaning for
the system under study, because we work in the domain very far
from applicability of the standard Debye-H\"{u}ckel theory. In
particular, the average number of ions in the volume $
\lambda_{D}^3 $ turns out smaller than unity. This is by no means
surprising, because there is no charge inversion in the
Debye-H\"{u}ckel theory, and to examine charge inversion we need
to go to the regime where this theory fails.

Since we do not have the $1:1$ salt in our simulation, we should keep
in mind that correlations between strongly charged $Z$-ions may be
significant even away from the macroion. Indeed, in the theory of
charge inversion \cite{Shklovskii}, the role of correlations is
emphasized for the $Z$-ions in the vicinity of the macroion, where
their concentration is particularly large. In our system, the
concentration of the $Z$-ions is not very small even in the bulk, and
we face the situations in which correlations between the $Z$-ions
away from the macroion affect our results. For such cases, we make
additional runs with the reduced concentration in the $Z:1$ salt.
However, we do not use this reduced concentration for all the runs,
because such system is more prone to noises and fluctuations, requiring
larger statistics to obtain reliable results.

Calculation of the Coulomb forces under the periodic boundary
conditions involves the charge sum in the first Brillouin zone
and their infinite mirror images (the Ewald sum\cite{Ewald}).
The sum is calculated with the use of the PPPM algorithm
\cite{Eastwood,Deserno}.
We use $ (32)^{3} $ spatial meshes for the calculation of the
reciprocal space contributions to the Coulomb forces, with
the Ewald parameter
$ \alpha \approx 0.262 $ and the real-space cutoff $ r_{cut}= R_{i} +
10a $, where $ R_{i} $ is the radius of the $ i $-th ion.
A uniform electric field $ E $ is applied in the $ x $-direction.

When starting the molecular dynamics simulation run, we prepare
an initial state by randomly positioning all the ions and neutral
particles in the simulation domain and giving Maxwell-distributed
random velocities corresponding to the temperature $ T_{\rm initial} $.
We integrate the Newton equations of motion with the use of the
leapfrog method \cite{textMD}, which is equivalent to the Verlet
algorithm.
In the absence of the electric field ($ E=0 $), our system
is closed, and its energy is conserved.
After an initial transient phase, the distribution of velocities
relaxes to a Maxwellian, corresponding to an equilibrium sampling of
the microcanonical ensemble.
This new Maxwell distribution has the temperature $ T $,
which is a little higher than $ T_{\rm initial} $, because of the
release of the potential energy due to screening, i.e.,
local balancing of charges.
We adjust $ T_{\rm initial} $ such that $ k_B T = \varepsilon $.
This makes $ \varepsilon $ to be the unique relevant scale of energy,
and, accordingly, we put $ \tau = a \sqrt{m / \varepsilon} $ as
the unit of time.
We choose $ \Delta t= 0.01 \tau $ as the integration time step.
The simulation runs are executed up to $ 1000 \tau $.

\subsection{Hydrodynamic interactions, their screening, and the
temperature control}
\label{Sec.2.2}

When an external electric field is present, it performs work on
the system. The corresponding energy, which is the Joule heat, is
transferred to background neutral particles through collisions
with accelerated ions. In our work, we simulate an
electrophoretic bath that is kept at a constant temperature $ T $.
For this purpose, we pretend that all neutral particles go through
the thermal bath of infinite heat capacitance, whenever they cross
the boundaries of the simulation domain (at the center of which
the macroion is located at every moment). Operationally,
we refresh the velocities of the neutral particles
according to the thermal distribution when they cross the domain
boundaries. This procedure maintains temperature stably to within
5\%.

Two closely related factors are potentially dangerous as they might
affect the molecular dynamics simulation results. One factor is the
finite-size of the simulation domain, and the other is the
long-range character of both hydrodynamic and Coulomb
interactions. These problems become particularly important
because some of the methods to simulate a constant temperature
thermal bath are believed to lead to the screening of hydrodynamic
interactions. This is obviously not acceptable in the simulation
where the long range character of hydrodynamic interactions is
expected to be important \cite{Dunweg}.

Following \cite{Ajdari,Viovy}, we argue that hydrodynamic
interactions are in fact effectively screened in our system and,
therefore, that the domain size in our simulation is quite sufficient
and the heat bath procedure described above is benign and
reliable.

To understand the situation, it is worth discussing the major
point - the screening of hydrodynamic interactions. To begin with,
why are hydrodynamic interactions long ranged? That fact can be
understood well from the point of view of momentum conservation.
Consider a particle immersed in a fluid and suppose that we pull
this particle with an externally applied force (such as gravity).
Obviously, this force transfers momentum into the system and,
however large the container may be, this momentum must be transported
away through the container walls. This necessitates the long range
character of the hydrodynamic field. More accurately, if we surround
our object by an arbitrarily large closed surface, then (under the
stationary conditions) the outflow of momentum through this
surface must be equal to the inflow of momentum due to the
external force. Because of the obvious analogy with the Gauss
theorem in electrostatics, we see that hydrodynamic field is just
as long ranged as the Coulomb field (even though it has a more
complicated vector structure).

The above description must be modified significantly when the
applied external force is due to the electric field and
the solution is neutral as a whole. In this case, there is no
overall inflow of momentum into the system, and therefore, there
should not be any outflow through the walls of the container. More
specifically, if there is one negatively charged macroion as in
our simulation, it is surrounded by positively charged counterions
and negatively charged coions such that the total charge of the
crowd effectively vanishes at some finite distance. In the simple
case of the Debye-H\"{u}ckel theory, this happens at about the Debye
screening length $ \lambda_{D} $ from the macroion surface. Therefore,
no momentum is transported further away, and hydrodynamic
interactions are screened at the distances of the order of
$ \lambda_{D} $ \cite{Ajdari,Viovy}.

In this paper, we treat a more complicated situation in which
the Debye-H\"{u}ckel theory does not apply and it is not easy to
judge {\em a priori} at which distances the hydrodynamic
interactions are screened. Nevertheless, since the system is
neutral, hydrodynamic interactions must be screened. We therefore
perform special test (described in Sec.\ref{Sec.3.2.1}) looking at
the dependence of the macroion drift on the simulation
domain size. We find that the drift is essentially
size-independent at $L > 20 a$ which is the direct manifestation
of the screening of hydrodynamic interactions.

Since hydrodynamic interactions are screened, our simulation is
not very sensitive to the method of maintaining the constant
temperature. To test it, we have performed separate runs at weak
electric fields, $ E \le 0.3 \varepsilon/ae $, in which case we
can run the simulation even without any heat drainage for a
significant period of time before any noticeable heating of the
system; the results of these control runs are within error bars of
the data obtained using the thermal bath (Fig.\ref{Fig.mu_qsurf}).

\section{Simulation Results}
\label{Sec.3}

\vspace*{-0.3cm}
\subsection{General Properties}

\begin{figure}
\resizebox{0.50\textwidth}{!}{%
\includegraphics{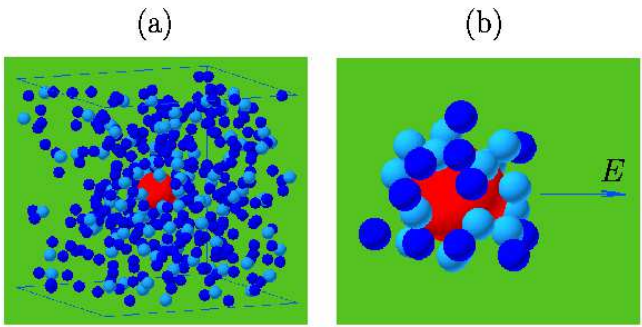}}

\caption{ Bird's-eye view of (a) all the ions
in the simulation domain and (b) the screening ion atmosphere
within $ 3a $ from the macroion surface. A macroion with charge $
Q_{0}= -30e $ and radius $ R_{0}= 3a $ is a large sphere in the
middle; counterions ($ Z= 3 $) and monovalent coions are shown by
light and dark blue spheres, respectively. The arrow to the right
shows the direction of the electric field ($ x $-axis), with $ E=
0.3 \varepsilon /ea $. }
\label{Fig.1}
\end{figure}

Our simulation results are shown in Figs.1-7. Fig. \ref{Fig.1} is
a bird's-eye view of (a) all the ions and (b) the vicinity of the
macroion. Counterions are shown in light blue and coions in dark
blue (neutral atoms are not shown). In this figure, the macroion
charge is taken to be $ Q_{0}= -30e $, its radius $ R_{0}= 3a $,
counterion valence $ Z= 3 $, and the electric field $ E= 0.3
\varepsilon /ea $. It is seen that the macroion is predominantly
covered by the counterions. As in the case without the electric
field \cite{Tanaka}, the radially integrated charge has a sharp
positive peak at a distance about $ a $ from the macroion surface.
This peak is due to the positive counterions being adsorbed on the
macroion surface. The value of the peak charge under the
conditions of Fig.\ref{Fig.1} is $ Q_{\rm peak} \approx 1.6
|Q_{0}| $.

\begin{figure}
\resizebox{0.45\textwidth}{!}{%
\includegraphics{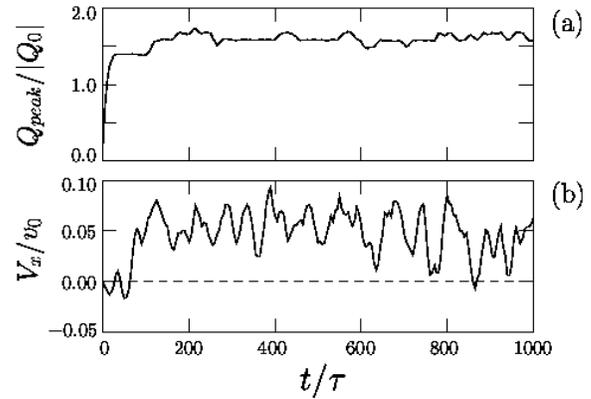}}

\caption{ Time history of (a) the {\it peak} charge $ Q_{peak} $
(defined as the maximum of radial charge distribution around the
macroion center) and (b) the macroion speed $ V_{x} $ normalized
by thermal velocity of neutral particles $v_{0}$. The macroion
complex drifts positively along the external electric field of $ E
> 0 $, which directly indicates the inversion of the charge sign.}
\label{Fig.t_hist}
\end{figure}

Fig. \ref{Fig.t_hist} demonstrates the time history of (a) the
"peak" charge and (b) the macroion drift speed for the parameters
of Fig.\ref{Fig.1}. At time $ t = 10 \tau $, we switch on the
external electric field. There is a short transient phase during
which a charge-inverted complex is formed through adsorption of
counterions to the macroion and condensation of coions on the
counterions. This process is reflected in a rather quick rise in $
Q_{\rm peak} $, as is shown in Fig.\ref{Fig.t_hist}(a). After the
transient phase, we observe a drift of the macroion in the
{\em positive} direction along the applied field. The fact that
the drift velocity is positive for the negative bare charge of the
macroion ($ Q_{0} < 0 $) is a direct manifestation of charge
inversion such that counterions are so strongly bound that they
pull the macroion with their motion.

Note that the drift velocity shown in Fig.\ref{Fig.t_hist}(b) is
small compared to the thermal velocity $ v_{0} $ of neutral
particles, $ \left< V_{x} \right> \sim 0.05 v_{0} $. Under this
condition, exchange of momentum between the macroion and neutral
particles is slow, and it requires many collisions (compare the
similar system in
Ref.\cite{LifshitzPitaevskiiPhysicalKineticsChaper2}). Therefore,
in terms of hydrodynamics, we are in the linear regime.
It means that friction force should be linear in the macroion velocity
and we expect the average drift speed to be given by the force
balance condition, $ Q^{*} E - \nu V_{x} = 0 $, where $ Q^{*} $ is
the effective net charge of the macroion complex and $\nu$ is the
hydrodynamic friction coefficient. We shall discuss later the
possibilities of determination of the effective net charge $ Q^{*}
$ based on this condition.

%We have measured this friction coefficient in an independent simulation,
%where we observe an exponential decay of the macroion velocity
%starting from $ 0.5 v_{0} $ for the case without an electric field.
%We find $ \nu \approx 9.3 m/ \tau $ for a {\it neutral}
%spherical macroion of radius $ R_{0} = 3a $.
%In reality, the relevant friction is expected to be larger
%because the macroion drifts by forming a complex with counterions
%and coions, which contributes to enhance its effective size.
%A measurement of the friction coefficient for a {\it fat} macroion
%with attached co/counterions in the ionic and neutral solvent
%environment, as is typical during our simulations,
%yields $ \nu \approx 18.2 m/\tau $.
%This implies that the effective radius of the macroion complex
%viewed in the electrolyte fluid Eq.(\ref{D_Stokes}) is
%$ R_{eff} \sim 4.2a = R_{0} + 1.2a $.
%The increase in the effective radius is a few Debye lengths.

Fig. \ref{Fig.t_hist} also shows significant temporal fluctuations
in the drift speed. Inspection reveals that they are larger than
what one expects for random kicks of neutral particles. These
fluctuations indicate that neither the counterions permanently
stick to fixed points on the macroion nor the coions attach to the
counterions, but that they are being replaced from time to time.
The fluctuations of this type are actually seen in a video movie.

\subsection{Parameter Dependences}
\label{Sec.3.2}

\vspace*{-0.3cm}
\subsubsection{Linear Regime}
\label{Sec.3.2.1}

\begin{figure}
\resizebox{0.45\textwidth}{!}{%
\includegraphics{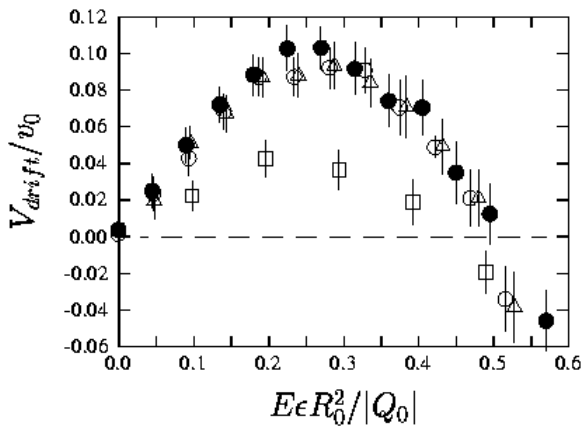}}

\caption{ Dependence of the macroion drift speed $
V_{\rm drift} $ (in the units of $ v_{0} $, the thermal speed of
neutral particles) on the electric field $ E $ for a macroion of
various radii and charges: $ R_{0}= 3a $ and $Q_{0} = -30 e$
(filled circles); $ R_{0}= 4a $ and $ Q_{0}= -50e $ (open
triangles); and $ R_{0}= 5a $ and $Q_{0} = -80e $ (open circles);
$R_{0} = 5 a$ and $Q_{0} = - 51 e$ (open squares). The valence of
counterions is $ Z= 3 $. }
\label{Fig.Vx_E}
\end{figure}

The dependence of the average macroion drift speed $ V_{\rm drift}
$ on the electric field is shown in Fig.\ref{Fig.Vx_E}. In this
figure, we show together the results of several runs,
corresponding to different values of the macroion charge $Q_0$ and
macroion radius $R_0$. First and foremost, the sign of the drift
velocity in moderate fields corresponds to the sign of inverted
charge. This is the central observation of our work. The figure
clearly demonstrates the overall pattern of the drift velocity
dependence on the applied field, beginning with the linear regime in a
weak field, followed eventually by a breakdown of charge inversion
in a strong field.

Let us discuss the linear drift regime for small electric fields $
E \le 0.2 |Q_{0}|/\epsilon R_{0}^{2} $, where $ V_{\rm drift} $
increases linearly with the field strength. This regime
corresponds to the usual Ohm's law, where the net charge of the
complex is insensitive to the strength of the electric field. A
macroion drifts together with its strongly adsorbed counterions as
a complex. That is, the electric field is not strong enough to
affect the binding of counterions to the macroion.

\begin{figure}
\resizebox{0.45\textwidth}{!}{%
\includegraphics{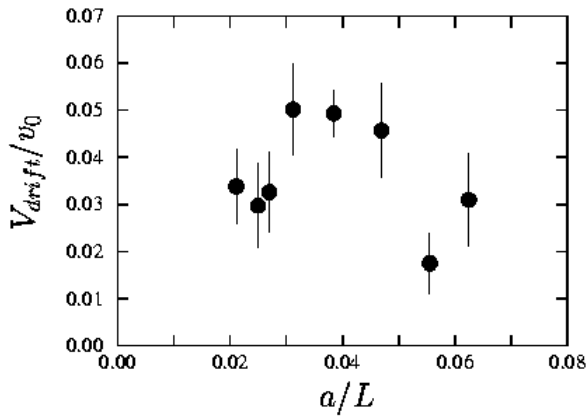}}

\caption{ Effect of the finite domain size $ L $
on the drift speed $ V_{drift} $ for a macroion with $ R_{0}= 3a
$, $ Q_{0}= -30e $, the electric field $ E= 0.3 \varepsilon /a e
$, and counterion valence $Z = 3$. The error bars correspond to
the root mean square velocity fluctuations as seen in Fig.
\protect\ref{Fig.t_hist}.}
\label{Fig.mu_L}
\end{figure}

At this stage, it is necessary to check the issue of
hydrodynamic interactions and their screening. For that purpose,
we show in Fig. \ref{Fig.mu_L} the effect of the simulation domain size
$ L $ on the macroion drift speed. By a series of the runs with
different domain sizes and under fixed number density of neutral
particles and ionic strength, we have confirmed that at $L = 32
a$, which is the domain size for the majority of our simulations,
the domain-size dependence is essentially leveled off. This is to
be contrasted with polymer chains \cite{Dunweg}, in which
case hydrodynamic interactions lead to the finite size effect
essentially linear in $a/L$.

We have also inspected the fluid flow of neutral particles around
the macroion. When rapid fluctuations are averaged out, this flow
field does not exhibit any patterns protruding away from the
macroion. This fact implies that the flow of neutral particles
induced by the motion of the macroion and other ions is screened
at short distances \cite{Ajdari,Viovy}.

The saturation of the $a/L$ dependence (Fig. \ref{Fig.mu_L}) and
the inspection of neutral particle flow patterns both confirm that
hydrodynamic interactions are screened in our system, thus making
reliable our simulation approach based on the finite domain and
the heat bath. The thermal bath at the distant boundaries does
not affect the measured macroion mobility.

The small electric field regime is characterized by the mobility,
$ \mu = \left< V_{x} \right> / E $. This quantity is plotted in
Fig. \ref{Fig.mu_qsurf} as a function of the macroion bare charge
$Q_{0}$, or, more specifically, on the surface charge density of
the macroion $ |Q_{0}|/R_{0}^{2} $.

\subsubsection{Can we measure net charge based on the mobility
measurement data?}

\begin{figure}
\resizebox{0.45\textwidth}{!}{%
\includegraphics{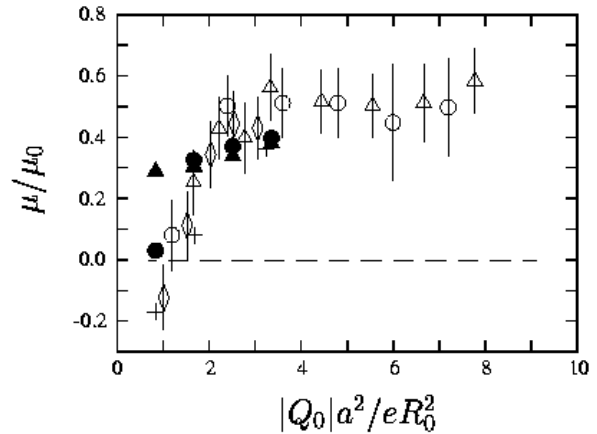}}

\caption{ 
Dependence of the macroion mobility $ \mu $ on the surface charge
density $ Q_{0}/R_{0}^{2} $ for the macroion radius $ R_{0}= 3a $
(open triangles), $ R_{0}= 5a $ (open circles), and $ R_{0}= 7a $
(open diamonds), where $ \mu_{0}= v_{0}/ (|Q_{0}^{(0)}|/\epsilon
(R_{0}^{(0)})^{2}) $ with $ Q_{0}^{(0)}= -30e $ and $ R_{0}^{(0)}=
3a $.  The valence of the counterions is $ Z= 3 $, the number 
of the $ Z:1 $ salt is $ N^{+}= (N^{-}e +|Q_{0}|)/Ze $ and 
$ N^{-}= 300 $, the electric field is $ E= 0.3 \varepsilon /ae $, 
and the temperature is $ e^2 /\epsilon ak_{B}T= 5 $. 
The filled circles and triangles show the cases with reduced number 
of the $ Z:1 $ salt such that $ N^{-}= $ 90 and 30, respectively.  
The crosses are the reference data obtained without the thermal bath 
for $ R_{0}= 5a $ and the $ Z:1 $ salt with $ N^{-}= 300 $.} 
\label{Fig.mu_qsurf}
\end{figure}

\begin{figure}
\resizebox{0.45\textwidth}{!}{%
\includegraphics{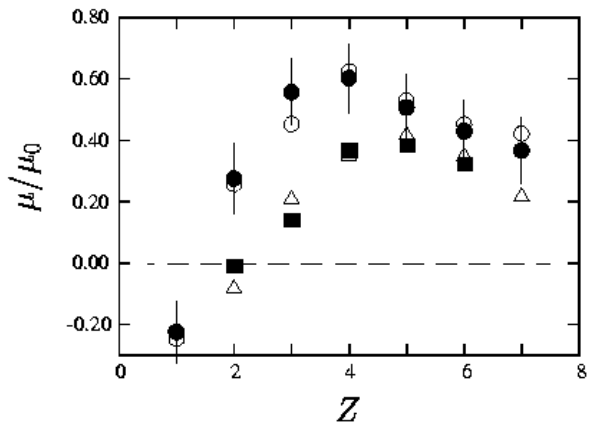}}

\caption{ Dependence of the macroion mobility $ \mu $ on the
valence of the counterions $ Z $ for the runs: 
$ R_{0}= 3a $ and $ Q_{0}= -30e $ (filled circles), $ R_{0}= 5a $
and $ Q_{0}= -80e $ (open circles), and $ R_{0}= 5a $ and $ Q_{0}=
-40e $ (filled squares).  Here, the number of the $ Z:1 $ salt
is $ N^{+}= (N^{-}e +|Q_{0}|)/Ze $ and $ N^{-} \approx 300 $, 
the external electric field is $ E= 0.3 \varepsilon /ae $, where 
$ \mu_{0}= v_{0}/ (|Q_{0}^{(0)}|/ \epsilon (R_{0}^{(0)})^{2} $ 
with $ Q_{0}^{(0)}= -30e $ and $ R_{0}^{(0)}= 3a $.
A series of the runs with reduced number of the $ Z:1 $ salt
$ N^{-} \approx 30 $, $ R_{0}= 5a $, and $ Q_{0}= -80e $ are 
shown with open triangles.}
\label{Fig.mu_Z}
\end{figure}

\begin{figure}
\resizebox{0.45\textwidth}{!}{%
\includegraphics{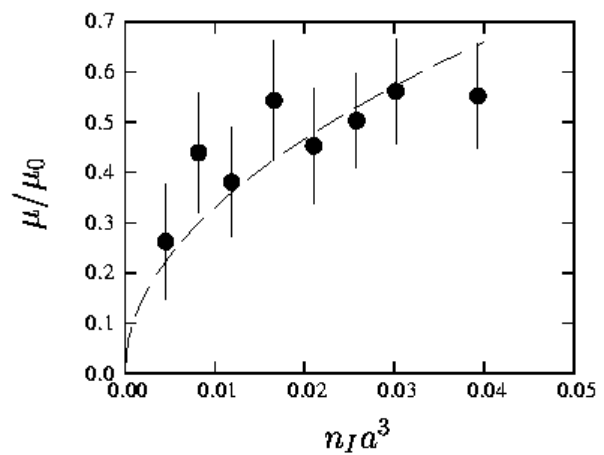}}

\caption{ Dependence of the macroion mobility $ \mu $ on the
salt ionic strength, $ n_{I}= (Z^{2} N^{+} +N^{-})/L^{3} $.
The parameters are $ Q_{0}= -30e $, $ R_{0}= 3a $,
$ E= 0.3 \varepsilon /ae $, $ Z= 3 $ and $ e^2 /\epsilon ak_{B}T=5 $,
which yield the Debye length $ \lambda_{D} \sim 0.45a $ for
$ n_{I}=0.04a^{-3} $. }
\label{Fig.mu_salt}
\end{figure}

Let us now discuss a practically important question: Can we
determine the effective net charge of the macroion complex $Q^{*}$
based on the data of mobility measurements, i.e., based on the
data of Fig.\ref{Fig.mu_qsurf} - Fig.\ref{Fig.mu_salt} ?
Physically, as we have already mentioned, the charge $ Q^{*} $
is determined by the force balance
condition $ Q^{*} E - \nu V_{x} = 0 $, or $Q^{*} = \mu \nu$.
Therefore, determination of the net charge requires knowledge of
both the friction coefficient $\nu$ and the mobility $\mu$.

Importantly, the friction coefficient cannot be determined by the
usual Stokes formula $\nu_S = 6 \pi \eta R$, where $\eta$ is the
solvent viscosity. The problem is that the real friction is
enhanced by the screening of hydrodynamic interactions
\cite{Ajdari,Viovy}. The Stokes formula is simply understood by
that the friction force in general should be proportional to $\eta
(v/A) B^2$, where $A$ is the length scale over which velocity
changes, $B^2$ is the relevant surface area. For the Stokes
problem, we have $A \sim B \sim R_{0} $. 
By contrast, when the drift is caused by the action
of the electric field on the overall neutral system, the distance
$A$ becomes much smaller. In the Debye-H\"uckel theory, it turns
out to be of the order of the Debye screening length, $ A \sim
\lambda_D $. In this case, the friction coefficient becomes
\begin{equation} \nu = \nu_S R_{0}/\lambda_D \ , \label{eq:friction}
\end{equation}
i.e., it is enhanced by the factor $ R_{0}/\lambda_D $ compared to
the usual Stokesian friction. Although we work under the
conditions where the Debye-H\"uckel theory is not applicable, and we 
do not know exactly which length should be there instead of
$\lambda_D$ in Eq.(\ref{eq:friction}), this length is clearly 
smaller than $R_{0}$ and independent of $R_{0}$. Therefore, the
friction coefficient is proportional to $R_{0}^2$ - unlike more
familiar Stokes case where it scales like $R_{0}$.

Under usual circumstances where the charge $ Q^{*} $ is known, the
friction coefficient scaling as $ R_{0}^2 $ implies that mobility
$\mu = Q^{*}/ \nu \propto Q^{*}/R_{0}^2$ is determined by the
{\it surface charge density}, not by the charge and the surface
area separately. This fact was known to M.Smoluchowski already
a century ago \cite{Smoluchowski}.
In our simulation, effective charge is not known {\em a priori},
and the logic needs to be reversed. Figure \ref{Fig.mu_qsurf}
indicates that mobility $ \mu $ is essentially a constant when the
macroion bare surface charge density is not too small ($ |Q_{0}|
a^2 /e R_{0}^2 \ge 3 $). Given that $\nu \propto R_{0}^2$, we
conclude that the effective bare charge $Q^{*} = \mu \nu$ is
proportional to the surface area of the macroion; namely, charge
inversion is characterized by the overcharging density.
This agrees with the theory \cite{NguGS,review}.

In order to perform at least very rough quantitative estimate of
charge $Q^{*}$ based on the mobility data, we need to know the
Stokesian friction coefficient $\nu_S$ (or equivalently, we need
to know the viscosity of our model solvent).  We measure it in a
separate molecular dynamics run, by observing an exponential decay
of the macroion velocity starting from $ 0.5v_{0} $ for the case
without an electric field. We find $ \nu_{S} \approx 9.3 m/\tau $
for a spherical particle of the radius $ R_{0}= 3a $ and $ \nu_{S}
\approx 18.2 m/\tau $ for the macroion complex with adsorbed
counterions and coions. These two estimates provide lower and
upper bounds for the charge $Q^{*}$. Assuming $ R_{0}/\lambda_{D}
\approx 6 $ and $ \mu \approx 0.5 \mu_{0} $ (saturation regime in
Fig.\ref{Fig.mu_qsurf}) yields the inverted charge $ Q^{*}$ between 
$ 7e $ and $ 20e $. This is in rough agreement with the $Q_{peak}$
measurements. 

Special attention must be paid to the small bare surface charge
density case, for which Fig.\ref{Fig.mu_qsurf} indicates the
decrease in the reversed mobility. For some cases, the reversed
mobility even disappears altogether, changing to normal,
non-reversed mobility, $ \mu <0 $ when the macroion bare surface
charge density decreases to about $ |Q_{0}| a^2 /eR_{0}^2 $
$ \sim 1 $. It turns out that this is the manifestation of correlations
between $Z$-ions in the bulk solution away from the macroion.
Indeed, when the macroion is only weakly charged, the correlations of
$Z$-ions in its vicinity are not much stronger than in the bulk,
which suppresses charge inversion.
Simple estimate shows that, under the conditions when $\mu$ gets
small or negative in Fig.\ref{Fig.mu_qsurf}, the "Wigner cell"
radius of the $Z$-ions on the macroion surface, $ R_{W} = 2R_{0}
(eZ/ |Q_{0}|)^{1/2} $ (which is about $3.4a $ for $ |Q_{0}| a^2
/eR_{0}^2 \approx 1 $ and $ Z= 3 $) becomes comparable to the
average spacing between the $Z$-ions in the bulk.
The inspection of the radial charge distribution functions around
the macroion for this case agrees with this interpretation.
It shows that the counterions are loosely bound to the macroion,
and that the coions form relatively strong bonds with the
counterions and drift together with them.

To examine the above interpretation further, we perform special
runs with reduced concentration of the $Z:1$ salt. The results of
these runs are shown in Fig.\ref{Fig.mu_qsurf} with filled circles
and triangles for the $ N^{+}= (N^{-}e +|Q_{0}|)/Ze $ $ Z $-ions 
with $ N^{-}= $ 90 and 30 negative coions, respectively.  
As anticipated, with fewer
$Z$-ions in the bulk, charge inversion is not interrupted even
at small macroion bare charges. We regard this a convincing proof
that charge inversion occurs at small concentration of the
$Z$-ions {\it despite} a larger entropy penalty.

Figure \ref{Fig.mu_qsurf} also shows with crosses the data
of the control run performed under the condition of the weak
electric field without any heat drainage (see Sec.\ref{Sec.2.2}).
These data are within error bars of the cases with the simulated
thermal bath.

%The net charge of the macroion complex is deduced from the
%mobility measured above in Fig.\ref{Fig.mu_qsurf}.
%For small surface charge density, one has
%\begin{eqnarray}
% & Q^{*} \sim Z (n|Q_{0}|)^{1/2} R_{0},
%\end{eqnarray}
%where $ Q^{*} \sim \nu \mu $ and the Debye-screening enhanced
%Stokes friction Eq.(\ref{D_Stokes}) have been used.
%For surface charge densities $ |Q_{0}|a^{2}/eR_{0}^{2} \ge 3 $,
%the mobility becomes nearly constant and insensitive to the
%spherical effect, as the Wigner radius is small in comparison with
%the macroion radius $ R_{W} < R_{0} $.
%We also note that the saturation level of the mobility increases
%with the valence for $ 2 \le Z \le 4 $ (this point will be further
%examined in Fig.\ref{Fig.mu_Z}).
%The constant mobility yields the net charge of the complex
%\begin{eqnarray}
% & Q^{*} \propto Ze R_{0}^{2},
%\end{eqnarray}
%which does not depend on the bare macroion charge $ Q_{0} $.
%This functional dependence agrees with the theoretical
%analysis for a plane geometry, $ Q_{th}^{*} \propto
%n_{I} Ze R_{0}^{2} $ in the $ \lambda_{D} \ll R_{W} $ regime
%\cite{NguGS,review}. This agreement confirms the screening of
%hydrodynamic interactions at the Debye length that occurs in the
%electrolyte solutions.

The dependence of the macroion mobility $ \mu $ on the valence of
the counterions $ Z $ in Fig.\ref{Fig.mu_Z} is physically
interesting, and also important for application purposes. For the
cases shown with filled and open circles, the surface charge
density of the macroion is chosen nearly the same $ |Q_{0}|/R_{0}^{2}
\sim 3 $ so that they reside in the saturation regime of
Fig.\ref{Fig.mu_qsurf}. We emphasize that the mobility for these
cases is given by the same curve. The mobility of the macroion is
{\em negative} when counterions are monovalent $ Z= 1 $, because
there is no charge inversion but only regular Debye screening.
Thus, the charge inversion phenomenon does not occur in the
solution of monovalent salt (provided that the co- and
counter-ions have the equal radius). For $ Z \ge 2 $, charge
inversion does take place, as manifested by the {\em positive}
mobility. These observations agree with a previous study for
planar charged surfaces \cite{Kjell}. A remarkable finding is that
the mobility here is maximized for the intermediate valence, $ Z
\approx 4 $, unlike the peak inverted charge that accounts for
static charge distribution of mainly counterions \cite{Tanaka}.

%The presence of an optimum valence for the reversed mobility in
%Fig.\ref{Fig.mu_Z} is explained by competition of two mechanisms.
%First, large valence is favorable for charge inversion because of
%strong Coulomb correlations among counterions.
%Second, on the other hand, the spherical effect due to large
%Wigner cell radius comparable to the macroion radius reduces
%such interactions.
%Large correlation holes on the macroion make very rugged
%(anisotropic) iso-potential surfaces that are produced by a
%macroion plus "surface" counterions as the valence becomes large.
%Thus, a fewer number of three-dimensional potential holes appear
%where counterions can be trapped stably.
%This mechanism also applies to the mobility for small surface
%charge densities in Fig.\ref{Fig.mu_qsurf}, where the ion correlation
%holes are again large compared to the macroion radius.

It is also noted in Fig.\ref{Fig.mu_Z} that, when the surface
charge density is reduced, both the magnitude of reversed
(positive) mobility and the range of $ Z $ where it occurs shrink
as shown by square symbols in the figure. The mobility for divalent
$Z$-ions is now negative. This corresponds to the small surface charge
density regime $ |Q_{0}|/R_{0}^{2} \sim 1.6 $ in Fig.\ref{Fig.mu_qsurf}.
Yet, the mobility is maximized for the intermediate valence,
$ Z \cong 5 $ in this case.
Also shown in Fig.\ref{Fig.mu_Z} are the results of the
control runs performed with reduced number of $Z$-ions (discussed
above in connection with Fig. \ref{Fig.mu_qsurf}). They again
reproduce the optimal valences for charge inversion.

Although somewhat speculative, we can try to apply the result of
Fig.\ref{Fig.mu_Z} to explain the electrophoretic mobility
measurements of nucleosome core particles in cation solution
\cite{Biomol}. What was observed is the increase in the
magnitude and range of the cation concentration for occurrence of
reversed mobility when spermidine salt (+3) was replaced with
spermine salt (+4), while charge reversal was not observed for any
concentration of Mg$ {}^{+2} $. It is worth stressing that
nucleosome particles is a complicated system in which many aspects
may be important. What we would like to say here is that our
results may be at least one of the factors relevant to the
experiments reported in the literature \cite{Biomol}.

The dependence of the macroion mobility on the salt ionic strength,
$ n_{I}= (Z^{2} N^{+} +N^{-})/L^{3} $ is shown in
Fig.\ref{Fig.mu_salt} for the counterion valence $ Z= 3 $ and
the temperature $ e^2 / \epsilon ak_{B}T= 5 $.
Here, the ionic strength is related to the Debye length
by $ \lambda_{D}= (\epsilon k_{B}T/8 \pi n_{I}e^{2})^{1/2} $.
The mobility increases quite rapidly for small ionic strength, and is
well fit by $ \mu \propto n_{I}^{1/2} $ as shown by a dashed curve.

It is a legitimate concern to ask whether the data of 
Fig.\ref{Fig.mu_salt} are affected by the correlations of $Z$-ions in the 
bulk which we discussed in connection with Fig. \ref{Fig.mu_qsurf}. 
The answer is negative; these data correspond to the saturation regime of 
Fig. \ref{Fig.mu_qsurf}. The distance between $Z$-ions in the bulk
drops to the value comparable to the Wigner-Seitz cell only when
$n_{I} a^3$ gets as large as $0.1$ or more.

The increase in the mobility with ionic strength contradicts the
intuition based on the Debye-H\"{u}ckel theory, and deserves a comment.
As it is explained in detail in \cite{review} and understood by a
number of authors cited there, charge
inversion itself grows with ionic strength. This happens because
charge inversion is the result of interplay between the repulsion of
counterions from each other and the attraction of each of them to
its own correlation hole. The latter occurs at a much shorter
distance than the former, and only the repulsion is strongly affected
by screening.
This is why the amount of charge inversion, hence the macroion
mobility, grows with increasing ionic strength.

It is also worth mentioning that the quick rise in the reversed
mobility at small ionic strength in Fig.\ref{Fig.mu_salt} agrees with
the colloidal mobility measurement for the case with trivalent salt
LaCl$ {}_{3} $ \cite{Elime,Gonza}.

\subsubsection{Nonlinear Regime}

Let us now return to Fig.\ref{Fig.Vx_E} to discuss the regime
that is nonlinear in the applied electric field.
As the figure indicates, the charge-inverted shell around the
macroion is destroyed for large electric fields.
Moreover, the critical field $ E_{c} $ at which this happens is
independent of the macroion size, which leads us to an estimate
\begin{equation}
E_{c} \approx 0.5 |Q_{0}|/ \epsilon R_{0}^{2} \ .
\label{critE}
\end{equation}
This result is quite interesting.
Indeed, $ |Q_{0}|/ \epsilon R_{0}^{2}$ is the electric field on
the macroion surface produced by the macroion bare charge.
Why does the critical field scale with the bare charge of the
macroion instead of the net charge of the complex ?
The reason is due to correlations between screening ions.
We noted while discussing Fig.\ref{Fig.t_hist} that the counterions
on the macroion surface are being replaced from time to time.
Consider how one $ Z $-ion can depart from the macroion surface.
Since this ion is surrounded by a correlation hole on the surface,
its departure requires work against the unscreened bare electric
field of the macroion as long as its distance from the surface is
smaller than the distance between the adsorbed $ Z $-ions.
Therefore, departure from the surface becomes possible when the
external field becomes comparable with this unscreened field;
the charge-inverted complex is no longer stable
at such a field strength.

The critical electric field in realistic situations is estimated
to be very large.
For the parameters $ R_{0} \approx 20 $\AA and $ Q_{0}
\approx 30 e $,
the critical electric field becomes as large as
$ E_{c} \approx 0.5 Q_{0} / \epsilon R_{0}^2
\approx 67 V/\mu m $, where we take into account the dielectric
constant of water $\epsilon \approx 80$ \cite{water}.
Although the critical field is large, it gives small energy to
the electric dipole of a water molecule, $ d \approx 2 \times 10^{-18}
{\rm esu} \cdot {\rm cm} $: $ E_{c} d/k_{B}T \sim 0.11 < 1 $.
This verifies the use of the model solvent of neutral particles
in the present molecular dynamics simulations.
In practice, the applied electric field is not expected to disrupt
the charge-inverted macroion complex.

\section{Summary}
\label{Sec.4}

In this paper, we performed molecular dynamics simulations with the use
of neutral-particle solvent, and measured the drift speed of a macroion
to obtain its mobility under electrophoresis in a multivalent $ Z:1 $
salt solution.

A weak electric field pulled the macroion complex in the direction
determined by the net inverted charge, instead of disrupting it.
The reversed mobility of the complex, $ \mu= V_{drift}/E $,
was shown to be nearly constant for the weak electric fields.
We showed the functional dependences of mobility in Figs.3, 5 and 6
of Sec.\ref{Sec.3}, respectively, against
the electric field strength $ E $, the surface charge density of
the macroion $ Q_{0}/R_{0}^{2} $, and the valence $ Z $ of the
counterions. The mobility was a function of the surface charge
density, instead of the bare charge and radius of the macroion
separately.
The reversed mobility increased rapidly with the salt ionic strength
$ n_{I} $ as $ \mu \propto n_{I}^{1/2} $.
%We established that finite domain-size effects due to hydrodynamic
%interactions only affected quantitative features of our results,
%not their qualitative essence.
Interestingly, the reversed mobility took a maximal value at the
intermediate valence of the counterions $ Z \cong 4 $.

%Also, there was a threshold of surface charge density for a
%spherical macroion to be charge inverted. These observations were
%attributed to the spherical effects due to large Wigner cell
%radius compared to the macroion radius. The reversed mobility
%became insensitive to the surface charge density for the surface
%charge density above $ |Q_{0}| a^{2} /eR_{0}^{2} \sim 3 $.

We confirmed the screening of hydrodynamic interactions at a
few Debye length. No specific flow patterns of neutral particles,
which one would expect for the sphere moving in a viscous fluid,
were detected around a macroion.

%We confirmed this screening through good agreement of the net
%charge of the macroion complex deduced from the simulation result
%and the theoretical prediction when the Debye-screened friction
%was applied.

In the large field regime, although academic because of its huge
value, electrophoresis was strongly nonlinear, and the field
stripped the screening counterions off the macroion.
The mobility of the macroion complex dropped significantly from
that of the linear regime, and the sign of the mobility flipped
back to non-reversed one above the critical electric field,
which was approximately half the macroion {\em unscreened} field.

In this study, we explicitly simulated the neutral particles of the
solvent to produce a reliable hydrodynamic background. On the
other hand, the limits of computational resources prevented us from
inclusion of the $1:1$ salt. The screening in our system was
exclusively accomplished by the $Z:1$ salt. For this reason, we
are not ready to make a quantitative comparison of our data with
the real experiments. The simulation including the $1:1$ salt is
currently under way.

{\small
The authors are grateful to Professor B.Shklovskii and Dr.T. Nguyen
for discussions, and to Dr.J.W.Van Dam for reading the manuscript.
One of the authors (M.T.) thanks Professor K.Kremer and Dr.C.Holm
for collaboration on the PPPM algorithm during his stay at the
Max-Planck Institut f\"ur Polymerforschung (Mainz, 1999)
under the support of the Max-Planck Society.
The numerical computation was performed with the vpp800/13
of the Institute of Space and Astronautical Science (Japan),
and partly with the Origin 3800 of the University of Minnesota
Supercomputing Institute.
}

\end{document}